\documentclass[runningheads]{llncs}
\usepackage{graphicx}
\usepackage{textcomp}
\usepackage{colortbl,xcolor}
\usepackage{float}
\usepackage{adjustbox}
\usepackage{hyperref}
\usepackage{amsmath}
\usepackage{booktabs}
\usepackage{hhline}

\usepackage{algorithm}
\usepackage[noend]{algpseudocode}
\makeatletter
\renewcommand{\ALG@beginalgorithmic}{\tiny}
\algrenewcommand\alglinenumber[1]{\tiny #1:}
\makeatother
\begin{document}
\title{Enhancing Data-Awareness of  Object-Centric Event Logs\thanks{This work was supported by the Fund for Scientific Research Flanders (project G079519N) and KU Leuven Internal Funds (project C14/19/082)}}
\titlerunning{Enhancing Data-Awareness of  Object-Centric Event Logs}
%
\author{Alexandre Goossens\inst{1}\orcidID{0000-0001-8907-330X},
Johannes De Smedt\inst{1} \orcidID{0000-0003-0389-0275},
Jan Vanthienen\inst{1}\orcidID{0000-0002-3867-7055}  \and
Wil van der Aalst \inst{2} \orcidID{0000-0002-0955-6940}}
%
%
\authorrunning{A. Goossens et al.}

\institute{Leuven Institute for Research on Information Systems (LIRIS), KU Leuven \email{\{FirstName\}.\{LastName\}@kuleuven.be}\\ \and Process and Data Science (PADS) chair, RWTH Aachen University \email{wvdaalst@pads.rwth-aachen.de}}
\maketitle              
\begin{abstract}
When multiple objects are involved in a process, there is an opportunity for processes to be discovered from different angles with new information that previously might not have been analyzed from a single object point of view.
This does require that all the information of event/object attributes and their values are stored within logs including attributes that have a list of values or attributes with values that change over time.
It also requires that attributes can unambiguously be linked to an object, an event or both. 
As such, object-centric event logs are an interesting development in process mining as they support the presence of multiple types of objects. First, this paper shows that the current object-centric event log formats do not support the aforementioned aspects to their full potential since the possibility to support dynamic object attributes (attributes with changing values) is not supported by existing formats.
Next, this paper introduces a novel enriched object-centric event log format tackling the aforementioned issues alongside an algorithm that automatically translates XES logs to this Data-aware OCEL (DOCEL) format.

\end{abstract}
\section{Introduction} 
In the last few years, object-centric event logs have been proposed as the next step forward in event log representation.
 The drive behind this is the fact that the eXtensible Event Stream (XES) standard \cite{xes2014ieee} with a single case notion does not allow capturing reality adequately  \cite{ghahfarokhi2020ocel}.
A more realistic assumption instead is to view a process as a sequence of events that interact with several objects.
Several object-centric event log representations have been proposed such as eXtensible Object-Centric (XOC) event logs \cite{li2018extracting}, Object-Centric Behavioral Constraint model (OCBC) \cite{artale2019modeling} , and most recently Object-Centric Event Logs (OCEL)\cite{ghahfarokhi2020ocel}.
The first two event log representations face scalability issues related to the storage of an object model with each event or to the duplication of attributes \cite{ghahfarokhi2020ocel}.
However, there is a difficult trade-off to be made between expressiveness and simplicity, leaving the recent OCEL  proposal as the most suitable for object-centric process mining as it strikes a good balance between storing objects, attributes and their relationships and yet keeping everything simple.

OCEL offers interesting new research opportunities not only for process mining with, e.g., object-centric Petri nets \cite{van2020discovering} or object-centric predictive analysis \cite{galanti2022object}, but also for decision mining \cite{hasic2017challenges}.
OCEL is already well on its way to become an established standard with a visualization tool \cite{ghahfarokhi2022python}, log sampling and filtering techniques \cite{berti2022filtering}, its own fitness and precision notions \cite{adams2021precision}, its own clustering technique \cite{ghahfarokhi2022clustering}, an approach to define cases and variants in object-centric event logs \cite{adams2022defining} and a method to extract OCEL logs from relational databases \cite{xiong2022extraction}. In this paper, attributes are considered to be logged together with events and objects in an event log and should relate clearly to their respective concepts, i.e., events, objects or both.  As such, OCEL could provide more analysis opportunities by supporting attributes having several values simultaneously, allowing attributes to change values over time and to unambiguously link attributes to objects, all of which is currently not fully supported but common in object-centric models such as  structural conceptual models like the Unified Modeling Language (UML)\cite{UML2017}.

For this purpose, this paper proposes an extension to OCEL called, Data-aware OCEL or DOCEL, which allows for such dynamic object attributes.
The findings are illustrated through a widely-used running example for object-centric processes indicating how this standard can also support the further development of object-centric decision/process mining and other domains such as Internet of Things (IoT) related business processes.
This paper also presents an algorithm to convert XES logs to DOCEL logs. Since many event logs are available in a "flat" XES format for every object involved in the process, not all information can be found in one event log. As such, providing an algorithm that merges these XES files into one DOCEL log would centralize all the information in one event log without compromising on the data flow aspects that make XES such an interesting event log format.

The structure of this paper is as follows: Section \ref{Motivation} explains the problem together with a running example applied on the standard OCEL form. Section \ref{A Proposed OCEL enrichment} introduces the proposed DOCEL format together with an algorithm to automatically convert XES log files into this novel DOCEL format.
Next, the limitations and future work of this work are discussed in Section \ref{Discussion}.
Finally, Section \ref{Conclusion} concludes this paper.

\section{Motivation} \label{Motivation}

The IEEE Task Force conducted a survey during the 2.0 XES workshop \footnote{https://icpmconference.org/2021/events/category/xes-workshop/list/?tribe-bar-date=2021-11-02} 
concluding that complex data structures, especially one-to-many or many-to-many object relationships, form a challenge for practitioners when pre-processing event logs.
By including multiple objects with their own attributes, object-centric event logs have the opportunity to address these challenges. 
This does entail that the correct attributes must be unambiguously linked to the correct object and/or activity to correctly discover the process of each object type as well as the relevant decision points \cite{van2020discovering}. The next subsection discusses the importance object attribute analysis had on single case notion event logs.

\subsection{Importance of object attributes in single case notion event logs}

Various single case notion process mining algorithms make use of both event and case attributes, e.g., in \cite{de2016general}, a framework is proposed to correlate, predict and cluster dynamic behavior using data-flow attributes. Both types of attributes are used to discover decision points and decision rules within a process in \cite{de2013data}. For predictive process monitoring, the authors of \cite{di2016clustering} develop a so-called clustering-based predictive process monitoring technique using both event and case data. Case attributes are also used to provide explanations of why a certain case prediction is made within the context of predictive process monitoring \cite{galanti2020explainable}.

The same challenges apply to decision mining which aims to discover the reasoning and structure of decisions that drive the process based on event logs \cite{vanthienen2021decisions}. In \cite{de2019holistic}, both event and case attributes are used to find attribute value shifts to discover a decision structure conforming to a control flow and in \cite{mannhardt2016decision}, these are used to discover overlapping decision rules in a business process.
Lastly, within an IoT context, it has been pointed out that contextualization is not always understood in a similar fashion as process mining does \cite{bertrand2022bridging}.
As such object-centric event logs offer an opportunity for these different views of contextualization to be better captured.

The previous paragraphs show (without aiming to provide an exhaustive overview) that various contributions made use of attributes that could be stored and used in a flexible manner. Unfortunately, as will be illustrated in the next subsections, the aforementioned aspects related to attribute analysis are currently not fully supported in object-centric event logs.

\subsection{Running Example} \label{Running Example}
\begin{figure*}[]
\centering
\includegraphics[width=1\linewidth]{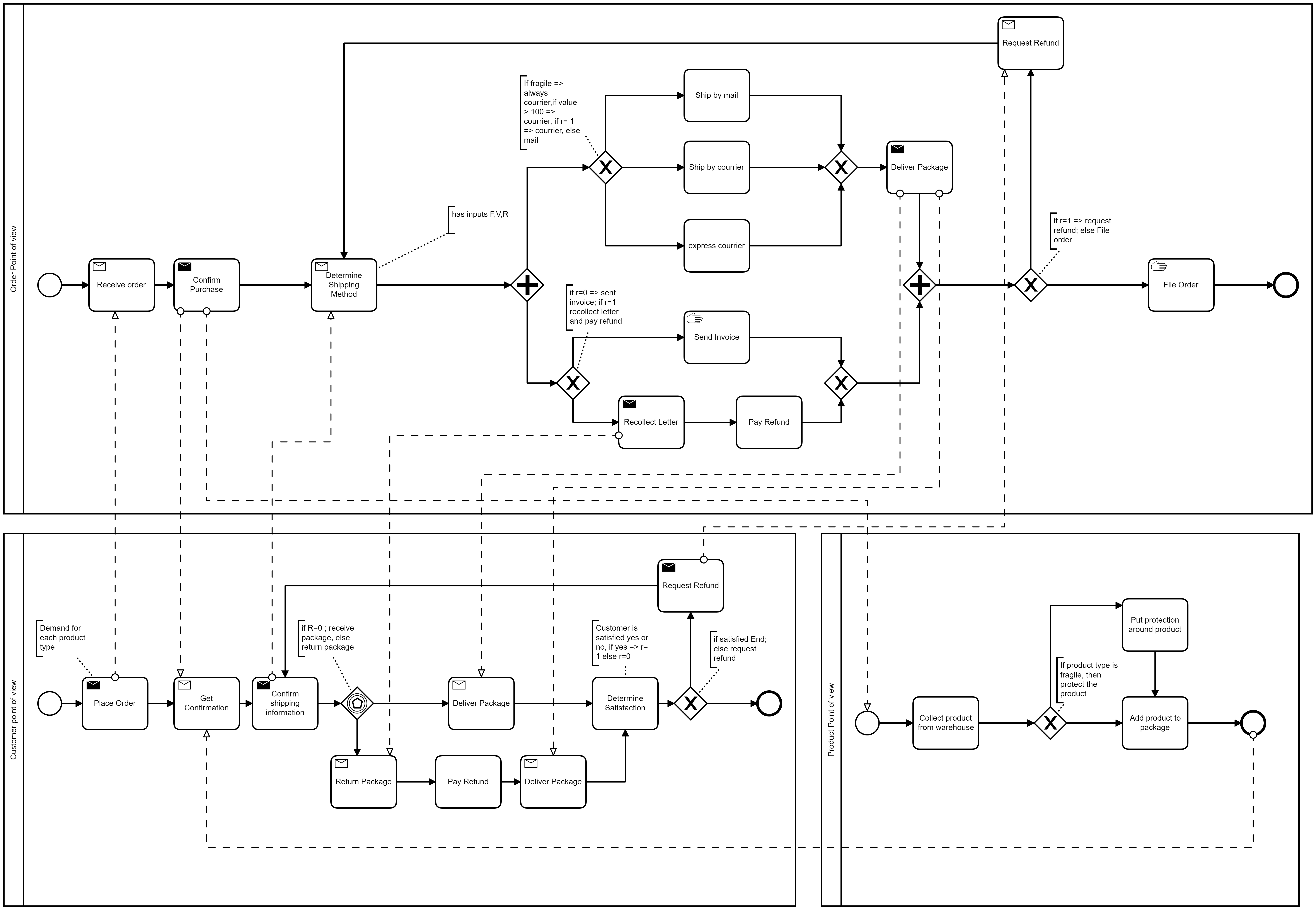}
\caption{BPMN model of running example} \label{BPMNrunning}
\end{figure*}

Consider the following adapted example inspired from \cite{de2019holistic} of a simple order-to-delivery process with three object types: Order, Product, Customer.
Figure \ref{BPMNrunning}\footnote{All figures are available in higher resolution using the following
\underline{\href{https://gitfront.io/r/user-6321558/g3WwhF8PAKKT/EdbA-ICPM2022/}{link}}.} visualizes the process. 

\textit{A customer places an order with the desired quantity for Product 1,2 or 3.
Next, the order is received and the order is confirmed.
This creates the value attribute of order.
Afterwards, the ordered products are collected from the warehouse.
If a product is a fragile product, it is first wrapped with cushioning material before being added to the package.
The process continues and then the shipping method needs to be determined.
This is dependent on the value of the order, on whether there is a fragile product and on whether the customer has asked for a refund.
If no refund is asked, this finalizes the process.
The refund can only be asked once the customer has received the order and requests a refund.
If that is the case, the order needs to be reshipped back and this finalizes the process.} 

\subsection{OCEL applied to the running example} \label{OCEL Representation of the running example}

In this subsection, the standard OCEL representation visualizes a snippet of this process. Table \ref{OCELevents} is an informal OCEL representation of events and Table \ref{OCELobjects} is an informal OCEL representation of objects. Figure \ref{OriginalUML} visualizes the meta-model of the original OCEL standard. Several \textbf{observations} can be made about the standard OCEL representation:

\begin{table*}[]
\caption{Informal representation of the events in an OCEL format}
\begin{adjustbox}{width=1\linewidth}
\begin{tabular}{|l|l|l|l|l|l|l|l|l|l|l|l|l|}
\hline
\textbf{ID} & \textbf{Activity}                 & \textbf{Timestamp} & \textbf{Customer} & \textbf{Order} & \textbf{Product Type} & \cellcolor{red!25} \textbf{Q1} & \cellcolor{red!25} \textbf{Q2} & \cellcolor{red!25} \textbf{Q3} & \cellcolor{red!25}  \cellcolor{red!25} \textbf{Refund} &  \cellcolor{red!25} \textbf{Order Value} & \cellcolor{green!25} \textbf{Resource} & \cellcolor{red!25} \textbf{Shipping Method} \\ \hline
\textit{e1}          & Place Order                       & 09:00              & \{\textit{c1}\}            & \{\textit{o1}\}         & \{\textit{p1,p2}\}             & 5           & 2           & 0           & 0               &                &                   &                          \\ \hline
\textit{e2}          & Receive Order                     & 10:00              &                   &  \{\textit{o1}\}         &                       &             &             &             &                 &                & Jan               &                          \\ \hline
\textit{e3}          & Confirm Purchase                  & 11:00              &                   &  \{\textit{o1}\}         &                       &             &             &             &                 & 95             & Jan               &                          \\ \hline
\textit{e4}          & Collect product from warehouse    & 12:00              &                   &  \{\textit{o1}\}        & \{\textit{p2}\}                &             &             &             &                 &                & Johannes          &                          \\ \hline
\textit{e5}          & Collect product from warehouse    & 12:00              &                   &  \{\textit{o1}\}          & \{\textit{p1}\}                &             &             &             &                 &                & Johannes          &                          \\ \hline
\textit{e6}          & Put protection around the product & 12:15              &                   &  \{\textit{o1}\}          & \{\textit{p1}\}                &             &             &             &                 &                & Johannes          &                          \\ \hline
\textit{e7}          & Add product to package            & 12:30              &                   & \{\textit{o1}\}       & \{\textit{p1}\}                &             &             &             &                 &                & Johannes          &                          \\ \hline
\textit{e8}          & Add product to package            & 12:30              &                   &  \{\textit{o1}\}      & \{\textit{p2}\}                &             &             &             &                 &                & Johannes          &                          \\ \hline
\end{tabular}
\end{adjustbox}
\label{OCELevents}
\end{table*}

\begin{table}[]
\centering
\caption{Informal representation of the objects in an OCEL format}
\begin{adjustbox}{width=0.75\linewidth}
\begin{tabular}{|l|l|l|l|l|l|l|l|l|l|l|l|l|}
\hline
\textbf{ID} & \textbf{Type} & \cellcolor{blue!25} \textbf{Name} & \cellcolor{blue!25} \textbf{Bank account} & \cellcolor{blue!25} \textbf{Value} & \cellcolor{blue!25} \textbf{Fragile} \\ \hline
\textit{c1}          & Customer      & Elien         & BE24 5248 54879 2659  &                &                  \\ \hline
\textit{o1}          & Order         &               &                       &                &                  \\ \hline
\textit{p1}          & Product       &               &                       & 15             & 1                \\ \hline
\textit{p2}          & Product       &               &                       & 10             & 0                \\ \hline
\textit{p3}          & Product       &               &                       & 20             & 1                \\ \hline
\end{tabular}
\end{adjustbox}
\label{OCELobjects}
\end{table}

\subsubsection{A: Attributes that are stored in the events table can not unambiguously be linked to an object.} 
The OCEL standard makes the assumption that attributes that are stored in the events table can only be linked to an event.
This assumption was taken for its clear choice of simplicity and it holds in this running example, which has straightforward attributes relationships and no changing product values over time.
Even though the given example is very obvious regarding how the attributes relate to the objects given the attribute names, this is not always the case.
If the value of a product could change over time, the product value attributes would have to be added to the events table but then there would be 4 attributes storing values, i.e., order value, product 1 value, product 2 value and product 3 value.
Knowing which attribute is linked to which object would then require domain knowledge as it is not explicitly made clear in the events table.
As such, this can be an issue in the future for generic OCEL process discovery or process conformance algorithms since prior to running such an algorithm, the user would have to specify how attributes and objects are related to one another.
    
\begin{figure*}[!]
\centering
\begin{minipage}{.45\textwidth}
  \centering
  \includegraphics[width=\linewidth]{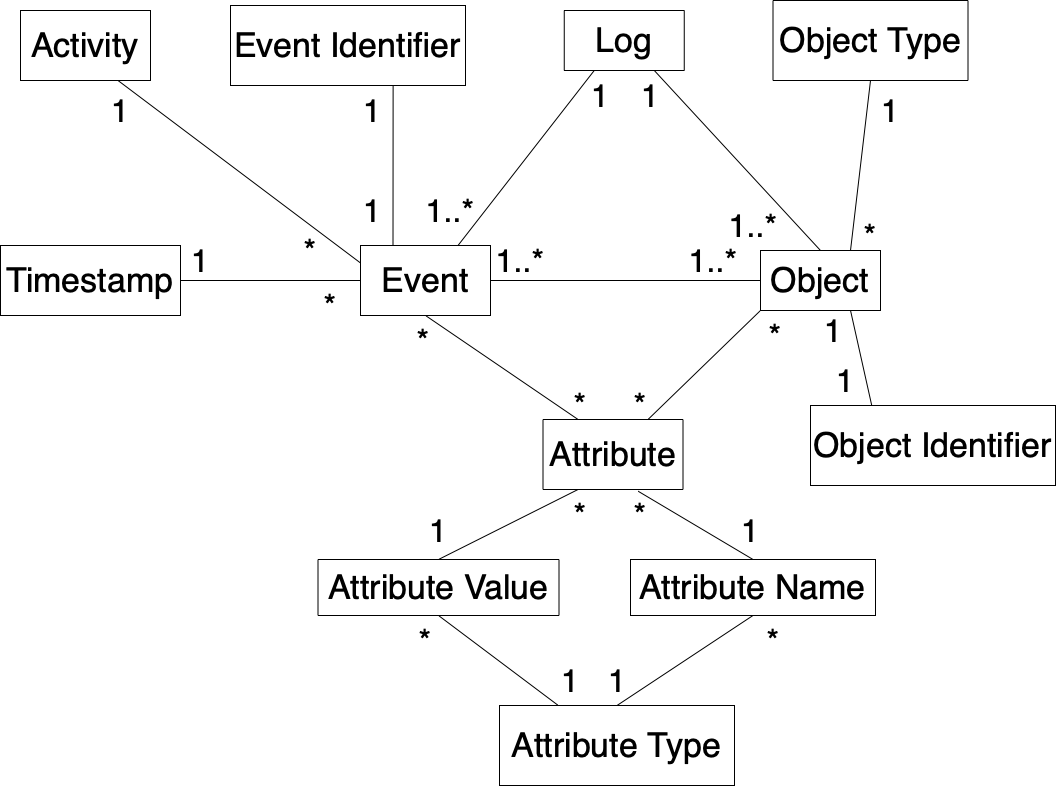}
 \caption{OCEL UML model from \cite{ghahfarokhi2020ocel}} 
  \label{OriginalUML}
\end{minipage}%
\begin{minipage}{.55\textwidth}
  \centering
  \includegraphics[width=\linewidth]{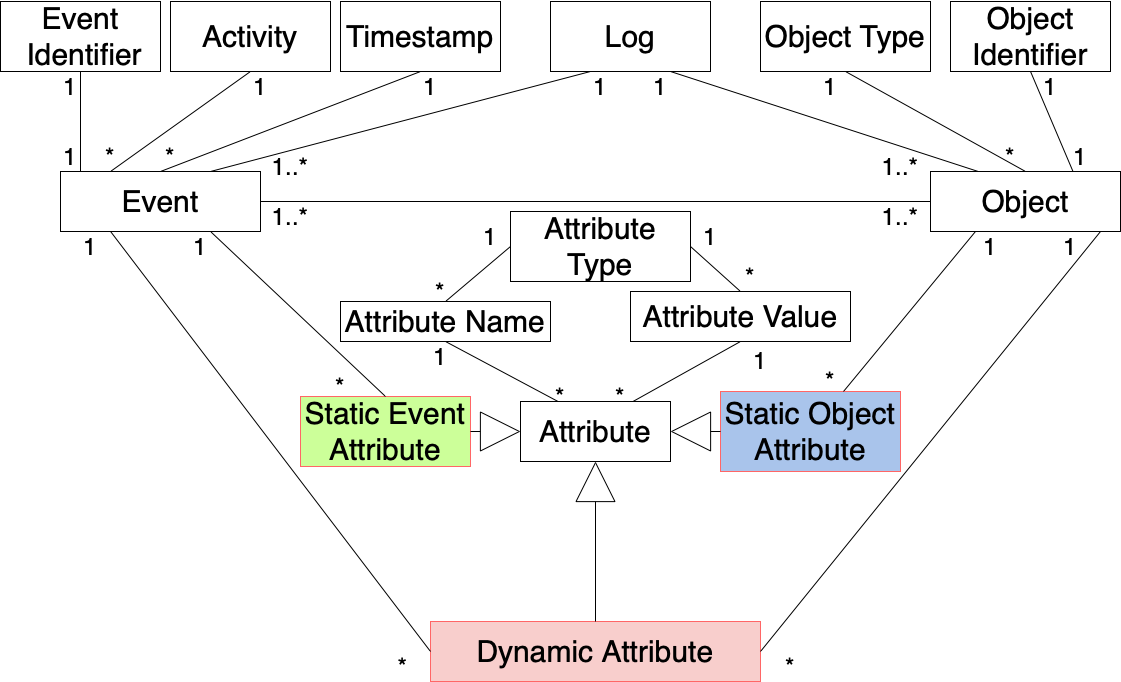}
  \caption{DOCEL UML model}
  \label{SuggestionUML}
\end{minipage}
\end{figure*}

\textbf{B: Based on the OCEL metamodel (Figure \ref{OriginalUML}), it is unclear whether attributes can only be linked to an event or an object individually or whether an attribute can be linked to both an event and an object simultaneously.} 
Since the OCEL standard did not intend for attribute values to be shared between events and objects by design to keep things compact and clear and since the OCEL UML model (Figure \ref{OriginalUML}) can not enforce the latter, Object-Constraint Language (OCL) constraints would have made things clearer.
Therefore, it might be beneficial to support the possibility \textit{to track an attribute change}, e.g., the \textit{refund} attribute of object \textit{Order} can change from 0 to 1 and back to 0 across the process.

\textbf{C: Attributes can only contain exactly one value at a time according to the OCEL metamodel (see Figure \ref{OriginalUML}).}
This observation entails two aspects.
First, it is unclear, based on the metamodel of Figure \ref{OriginalUML}, whether an attribute can contain a list of values.
It is not difficult to imagine situations with a list of values, e.g., customers with multiple bank accounts or emails, products can have more than one color.
Currently, OCEL supports multiple values by creating a separate column for each value in the object or event table.
This means that each value is treated as a distinct attribute , e.g., in the running example, a customer orders a quantity of product 1, 2 and 3.
This can be considered as 1 attribute with 3 values. However, in Table \ref{OCELevents}, the  columns Q1, Q2 and Q3 are considered to be separate attributes even though they could be considered as being from the same overarching attribute Quantity. 
Secondly, even if an attribute only has 1 value at a time, its value could change over time as well.
Such an attribute can be considered to have multiple values at different points in time.
    If a value were to change, currently, one would have to create a new object for each attribute change. Unfortunately, this only works to some degree since there are no object-to-object references (only through events) in the standard OCEL format. 
    Another possibility would require to unambiguously track the value of an attribute of an object to a certain event that created it.
    This is also valid within an IoT context with sensors having multiple measurements of the same attributes over time.
    As such, the first three observations clearly go hand in hand.

\textbf{D: Both the event and object tables seem to contain a lot of columns that are not always required for each event or object.} 
When looking at the events table, attribute \textit{Order Value} is only filled once with event ‘confirm purchase’ when it is set for order 1. One could either duplicate this value for all the next events dealing with order 1 or one could simply keep it empty. Therefore, in a big event log with multiple traces one could expect a lot of zero padding or duplication of values across events. Even though this issue is not necessarily present in a storage format, it still shows that ambiguity about attribute relationships might lead to wrongly stored attributes without domain knowledge.

\section{Data-aware OCEL (DOCEL)} \label{A Proposed OCEL enrichment}

Subsection \ref{DOCEL UML metamodel} introduces the DOCEL UML metamodel. Next, Subsection \ref{Data-aware OCEL proposal of the running example} applies DOCEL to the running example. Finally, Subsection \ref{Automatically converting XES log to DOCEL logs} introduces an algorithm to convert a set of XES files into this DOCEL format.

\subsection{DOCEL UML metamodel}\label{DOCEL UML metamodel}
To formally introduce the DOCEL standard, a UML class diagram is modeled (Figure  \ref{SuggestionUML}). UML diagrams clearly formalize how all concepts relate to one another in OCEL or DOCEL. Based on the observations from Section \ref{OCEL Representation of the running example}, the key differences with the UML class diagram of OCEL (Figure \ref{OriginalUML}) are indicated in color in Figure \ref{SuggestionUML} to enrich OCEL even further:

\textbf{1: Attribute values can be changed and these changes can be tracked.} 
By allowing ambiguities, domain knowledge becomes indispensable to make sensible and logical conclusions.
In the DOCEL UML model, attributes are considered to be an assignment of a value to an attribute name in a particular context event and/or object. A distinction is made between static and dynamic attributes.
Static event attributes and static object attributes are assumed to be linked to an event or an object respectively and only contain fixed value(s).
Static attributes are stored in a similar fashion as with the standard OCEL format, namely in the event or the object table, except that now each object type has an individual table to avoid having null values for irrelevant columns.
On the other hand, dynamic attributes are assumed to have changing values over time.
Dynamic attributes are linked to both an object and an event so that a value change of an attribute can easily be tracked.
Another design choice would be to store a timestamp with the attribute value instead of linking it to the event, however, this might lead to ambiguity in case two events happened at the exact same moment. As such, this proposal tackles observation \textbf{A}.
     
\textbf{2: Event attributes can unambiguously be linked to an object.}
This issue goes hand in hand with the previous proposal and is solved at the same time.
By distinguishing between dynamic and static attributes all relations between attributes, events and objects are made clear and ambiguities have been reduced.
A static attribute is either linked to an object or an event and its value(s) can not change over time.
A dynamic attribute is clearly linked to the relevant object and to the event that updated its value. 
The DOCEL UML model (Figure \ref{SuggestionUML}) can enforce that a static attribute must be linked with at least 1 event or at least 1 object since a distinction is made between static event attributes and static object attributes.
For dynamic attributes, this issue does not apply since it needs to both connected to both an object and an event anyhow. This proposal solves both observations \textbf{A \& B}.
    
\textbf{3: Attributes can contain a list of values.}
    Even though not all attributes have a list of values,
    supporting this certainly reflects the reality that multiple values do occur in organizations. In the DOCEL UML model (Figure \ref{SuggestionUML}) the 1 cardinality for Attribute Value allows both dynamic and static attributes to have complex values, e.g., lists, sets and records containing multiple values.
    In practice, these values are stored in the relevant attribute tables with a list of values. This proposal solves observation \textbf{C}.

\subsection{DOCEL applied to the running example} \label{Data-aware OCEL proposal of the running example}

Table \ref{dELevents} is the events table containing all the events together with their \textbf{static event attributes} (in green) in this case \textit{Resource}.
Complying with the DOCEL UML model, only static event attributes are found in this table which are solely linked to events. The main changes from the OCEL to the DOCEL tables have been  highlighted using the same color scheme as in the DOCEL UML model to show where the columns have been moved to in the DOCEL tables.

\begin{table*}[]
\caption{Informal representation of events with static attributes in a DOCEL format} \label{dELevents}
\begin{adjustbox}{width=\linewidth}
\begin{tabular}{|l|l|l|l|l|l|l|}
\hline
\textbf{EID} & \textbf{Activity}                 & \textbf{Timestamp} & \textbf{Customer} & \textbf{Order} & \textbf{Product Type} & \cellcolor{green!25}  \textbf{Resource} \\ \hline
\textit{e1}           & Place Order                       & 1/01/22 09:00      & \{\textit{c1}\}            & \{\textit{o1}\}         & \{\textit{p1,p2}\}             &                   \\ \hline
\textit{e2}           & Receive Order                     & 1/01/22 10:00      & \{\textit{c1}\}          & \{\textit{o1}\}         & \{\textit{p1,p2}\}             & Jan               \\ \hline
\textit{e3}          & Confirm Purchase                  & 1/01/22 11:00      &                   & \{\textit{o1}\}         & \{\textit{p1,p2}\}             & Jan               \\ \hline
\textit{e4}             & Collect product from warehouse    & 1/01/22 12:00      &                   & \{\textit{o1}\}        & \{\textit{p2}\}                & Johannes          \\ \hline
\textit{e5 }            & Collect product from warehouse    & 1/01/22 12:00      &                   & \{\textit{o1}\}       & \{\textit{p1}\}                & Johannes          \\ \hline
\textit{e6 }            & Put protection around the product & 1/01/22 12:15      &                   & \{\textit{o1}\}         & \{\textit{p1}\}                & Johannes          \\ \hline
\textit{e7  }           & Add product to package            & 1/01/22 12:30      &                   & \{\textit{o1}\}         & \{\textit{p1}\}                & Johannes          \\ \hline
\textit{e8 }            & Add product to package            & 1/01/22 12:30      &                   & \{\textit{o1}\}        & \{\textit{p2}\}                & Johannes          \\ \hline
\end{tabular}
\end{adjustbox}
\end{table*}

Tables \ref{Product table}, \ref{Order table}, \ref{Customer table} represent object type tables where the objects are stored.
Each object is given an object ID.
In this data-aware format, aligned with the UML model, a distinction is made between static attributes and dynamic attributes.
Static attributes are assumed to be immutable and, therefore, the \textbf{static object attributes} (in blue) are stored together with the objects themselves, e.g., \textit{customer name}, \textit{product value}, \textit{fragile} and \textit{bank account}. Notice how here, once again, the attributes can be clearly linked to an object. Table \ref{Order table} only contains primary keys because its attributes are dynamic attributes in this example.

\begin{table}[]
    \begin{minipage}{.25\linewidth}
      \centering
        \caption{Product Type table} \label{Product table}
        \begin{tabular}{|l|ll}

        \cline{1-1}
        \textbf{Products} &                                     &                                       \\ \hline
        \textbf{PID}      & \multicolumn{1}{l|}{\cellcolor{blue!25}\textbf{Value}} & \multicolumn{1}{l|}{\cellcolor{blue!25}\textbf{Fragile}} \\ \hline
        \textit{p1}                & \multicolumn{1}{l|}{15}             & \multicolumn{1}{l|}{1}                \\ \hline
        \end{tabular}
    \end{minipage} 
         \begin{minipage}{\linewidth}
     \caption{Order table} \label{Order table}
     \centering
       \begin{tabular}{|l|}

        \cline{1-1}
        \textbf{Orders}  \\ \hline
        \textbf{OrderID} \\ \hline
        \textit{o1}               \\ \hline
        \end{tabular}
        \end{minipage}%
        \\
    \begin{minipage}{\linewidth}
        \caption{Customer table} \label{Customer table}
        \centering
         \begin{tabular}{|l|ll}
        \cline{1-1}
        \textbf{Customer} &                                    &                                           \\ \hline
        \textbf{CID}       & \multicolumn{1}{l|}{\cellcolor{blue!25}\textbf{Name}} & \multicolumn{1}{l|}{\cellcolor{blue!25}\textbf{Bank account}} \\ \hline
        \textit{c1}                & \multicolumn{1}{l|}{Elien}         & \multicolumn{1}{l|}{BE24 5248 5487 2659}  \\ \hline
        \end{tabular}
    \end{minipage} 
   
\end{table}

The red tables \ref{Quantity table}, \ref{Value table}, \ref{Refund table}, \ref{Shipping method table} are \textbf{dynamic attribute} tables.
Dynamic attributes are assumed to be mutable and its values can change over time.
Using two foreign keys (event ID and object ID), the attribute and its value can be traced back to the relevant object 
as well as the event that created it.
Each attribute value is given an attribute value ID with the value(s) being stated in the following column.
This complies with the proposed UML model in Figure \ref{SuggestionUML} where dynamic attributes are clearly linked to the relevant event and relevant object.

\begin{table}[]
    \begin{minipage}{0.25\linewidth}
        \caption{Quantity table} \label{Quantity table}
        \begin{tabular}{|l|lll}
            \hhline{-~~}
            \cellcolor{red!25} \textbf{Quantity} &                                  &                                  &                                                      \\ \hline
            \textbf{QID}      & \multicolumn{1}{l|}{\textbf{Quantity}} & \multicolumn{1}{l|}{\textbf{EID}} & \multicolumn{1}{l|}{\textbf{OID}} \\ \hline
            \textit{q1}                & \multicolumn{1}{l|}{\{5,2,0\}}           &  \multicolumn{1}{l|}{\textit{e1}}           & \multicolumn{1}{l|}{\textit{o1}}           \\ \hline
        \end{tabular}
    \end{minipage} 
        \begin{minipage}{\linewidth}
      \caption{Order Value table} \label{Value table}
      \centering
        \begin{tabular}{|l|lll}
            \hhline{-~~~}
            \cellcolor{red!25} \textbf{Order Value} & \textbf{}                           & \textbf{}                         & \textbf{}                                                         \\ \hline
            \textbf{VID}   & \multicolumn{1}{l|}{\textbf{Value}} & \multicolumn{1}{l|}{\textbf{EID}} & \multicolumn{1}{l|}{\textbf{OID}}  \\ \hline
            \textit{v1}             & \multicolumn{1}{l|}{95}             & \multicolumn{1}{l|}{\textit{e3}}           & \multicolumn{1}{l|}{\textit{o1}}                    \\ \hline
        \end{tabular}
    \end{minipage}%

    \begin{minipage}{0.3\linewidth}
      \centering
        \caption{Refund table} \label{Refund table}
        \begin{tabular}{|l|lll}
            \hhline{-~~~}
            \cellcolor{red!25} \textbf{Refund} & \textbf{}                           & \textbf{}                         & \textbf{}                         \\ \hline
            \textbf{RID}    & \multicolumn{1}{l|}{\textbf{Refund  Value}} & \multicolumn{1}{l|}{\textbf{EID}} & \multicolumn{1}{l|}{\textbf{OID}} \\ \hline
            \textit{r1}              & \multicolumn{1}{l|}{0}              & \multicolumn{1}{l|}{\textit{e1}}           & \multicolumn{1}{l|}{\textit{o1}}           \\ \hline
            \textit{r2}              & \multicolumn{1}{l|}{1}              & \multicolumn{1}{l|}{\textit{e15}}          & \multicolumn{1}{l|}{\textit{o1}}           \\ \hline
            \textit{r3}              & \multicolumn{1}{l|}{0}              & \multicolumn{1}{l|}{\textit{e24}}          & \multicolumn{1}{l|}{\textit{o1}}       \\ \hline
        \end{tabular}
    \end{minipage} 
     \begin{minipage}{0.85\linewidth}
      \centering
        \caption{Shipping method table} \label{Shipping method table}
        \begin{tabular}{|l|lll}
            \hhline{-~~~}
            \cellcolor{red!25} \textbf{Shipping method} &                                       &                                   &                                   \\ \hline
            \textbf{SID}             & \multicolumn{1}{l|}{\textbf{Method}}  & \multicolumn{1}{l|}{\textbf{EID}} & \multicolumn{1}{l|}{\textbf{OID}} \\ \hline
            \textit{s1}                       & \multicolumn{1}{l|}{courrier}         & \multicolumn{1}{l|}{\textit{e11}}          & \multicolumn{1}{l|}{\textit{o1}}           \\ \hline
            \textit{s2}                       & \multicolumn{1}{l|}{express courrier} & \multicolumn{1}{l|}{\textit{e18}}          & \multicolumn{1}{l|}{\textit{o1}}         \\ \hline
        \end{tabular}
    \end{minipage} 
\end{table}

From the DOCEL log, the following things are observed:

\textbf{Attributes can unambiguously be linked to an object, to an event or to both an event and an object with the use of foreign keys.}

\textbf{Attributes can have different values over time,}with value changes directly tracked in the dynamic attributes tables. 
This means one knows when the attribute was created and for how long it was valid, e.g., refund was initialized to 0 by event 1, then event 15 set it to 1 and finally event 24 sets it back to 0. 
     
\textbf{Static and dynamic attributes can contain a list of  values} in the relevant attributes table, e.g., attribute Quantity.

\textbf{The amount of information stored has only increased with foreign keys.} Previously, the dynamic attributes would have been stored anyhow in the events table with the unfortunate side-effect of not being explicitly linked to the relevant object and with more columns in the events table. This essentially is a normalization of an OCEL data format. Even though it starts resembling a relational database structure, it was decided for this DOCEL format  to not include relations between objects.
Deciding on whether to include object models within event logs is essentially a difficult trade-off between complexity/scalability and available information within the event log.
From this perspective, the design choice of XOC and OCBC was mostly focused on reducing complexity \cite{ghahfarokhi2020ocel}, where we aim for an event log format that offers more information in exchange of a slightly increased complexity. As such, the DOCEL standard has decreased the amount of columns per table and thus observation \textbf{D} is solved as well.

\subsection{Automatically converting XES logs to DOCEL logs} \label{Automatically converting XES log to DOCEL logs}
Currently, research is focused on automatically converting XES logs to OCEL logs with a first proposal introduced in \cite{rebmannuncovering}.
Automatically transforming XES logs or an OCEL log to the proposed DOCEL log 
would mainly require domain knowledge to correctly link all attributes to the right object, but this is also required for a normal process analysis of an OCEL log.
Our algorithm can be found in Algorithm \ref{algXESOCEL}. 
This algorithm takes as input a set of XES files describing the same process and assumes that each XES file describes the process from the point of view of one object type.
The main ideas of the algorithm are as follows:
\begin{itemize}
    \item Line 3 starts the algorithm by looping over all XES-logs.
    \item Lines 4-8 create the object type tables with all their objects and static object attributes. In line 7, it is assumed that the trace attributes are not changing and solely linked to one object. Since the assumption is made that an XES file only contains one object type, these trace attributes can be considered as static object attributes belonging to that object.
    \item Lines 10-12 require the user to identify the static event attributes and the other event attributes that can be linked to an object. Next, a new EventID is made to know from which log an event comes from.
    \item In line 15, the dynamic attributes tables are constructed under the assumption that attributes that have not yet been identified as static object attributes or static event attributes are dynamic attributes.
    \item Lines 17-18 create the new chronologically ordered events Table $E$.
    \item Line 20 matches the events with the relevant objects based on the dynamic attributes tables using the new EventID. It should definitely also include the object related to the original traceID related to that event.
    \item Finally, lines 21-22 will create the final DOCEL eventIDs and update the eventID across all dynamic attribute tables.
\end{itemize}

\begin{algorithm}
\caption{ Algorithm to go from XES logs to DOCEL logs}\label{algXESOCEL}
\begin{algorithmic}[1]
\State $L \gets l$ \Comment{List of XES logs ($l$)}
\State $OT \gets ot$ \Comment{List of present object types}
\For{$l \in L$}
    \For{$ot \in (OT \in l$)}
    \State	 \textbf{Create} empty object type table
    \For{$o \in ot $} \Comment{Find all objects of an object type}
        \State  \textbf{Create} row with objectID and trace attributes    \Comment{Trace attributes = static object attributes}
    \EndFor

    \EndFor
    \For{$e \in L$}
    \State 	\textbf{Match} event attributes to the event or to an object
    \State \textbf{Create} $new EventID$ with log identifier \Comment{To distinguish similar events of different logs}
    \EndFor
    \State 	 \textbf{Create} event table $e_l$ with static event attributes.
    \State 	 \textbf{Create} dynamic attributes table with valueID, value(s) and two foreign keys \{$new EventID, object ID$\}
\EndFor
\State 	 \textbf{Create} empty event table $E$ with a column for every object type.
\State \textbf{Merge} all $e_l$ tables chronologically in $E$.

\For{$e \in E$}

\State \textbf{Find and insert} all objects related to $e$ in the relevant object type column 
\State \textbf{Create} unique DOCELeventID
\State \textbf{Update} all foreign keys of linked dynamic attributes with new DOCELeventID 
\EndFor
\end{algorithmic}
\end{algorithm}

\section{Limitations and Future Work} \label{Discussion}

To better store information about attributes, DOCEL comes with a variable number of tables.
However, the tables should be smaller as there are fewer columns compared to the standard OCEL format.  
It is still possible to only use certain attributes or attribute values for analysis by extracting the relevant attributes/values.
Instead of selecting a subset of columns with OCEL, the user selects a subset of tables in DOCEL which offer more information. Next, neither OCEL or DOCEL include the specific roles of objects of the same object type in an event, in case of a \textit{Send Message} event from person 1 to person 2, making it currently impossible to distinguish between the sender and the receiver.




To further validate the DOCEL format, the authors are planning to develop a first artificial event log together with a complete formalization of the DOCEL UML  with OCL constraints. Furthermore, directly extracting DOCEL logs from SAP is also planned. Regarding the algorithm to automatically convert XES logs to DOCEL logs, the authors are planning to extend the algorithm with a solution to automatically discover which attributes are linked to objects or events. Secondly, an extension to create a DOCEL log based on a single XES file with multiple objects is also planned. DOCEL however offers many other research opportunities such as novel algorithms for object-centric process discovery, conformance checking or enhancements which would further validate or improve the DOCEL format. Also other domains such as IoT-related process mining can be interesting fields to apply DOCEL on.

\section{Conclusion} \label{Conclusion}
This paper illustrates that the OCEL standard has certain limitations regarding attribute analysis, such as unambiguously linking attributes to both an event and an object or not being able to track attribute value changes.
To deal with these challenges, an enhanced Data-aware OCEL (DOCEL) is proposed together with an algorithm to adapt XES logs into the DOCEL log format.
With DOCEL, the authors hope that new contributions will also take into account this data-flow perspective not only for object-centric process and decision mining algorithms but also for other domains such as IoT-oriented process analysis. 

\bibliographystyle{splncs04}
\bibliography{main}

\end{document}